\documentclass[a4paper,preprint,aps]{revtex4}

\usepackage{amsmath}
\usepackage{amsfonts}
\usepackage{amssymb}
\usepackage{graphicx}
\usepackage{hyperref}

\begin{document}
\title{Diseases spreading through individual based models with realistic mobility patterns}
\author{A. D. Medus}\email{admedus@df.uba.ar}\affiliation{Departamento de F\'isica, Facultad de Ciencias Exactas y Naturales,
 Universidad de Buenos Aires, Pabell\'on 1, Ciudad Universitaria, Ciudad Aut\'onoma de Buenos Aires (1428), Argentina and IFIBA - CONICET}

\author{C. O. Dorso}\email{codorso@df.uba.ar}\affiliation{Departamento de F\'isica, Facultad de Ciencias Exactas y Naturales,
 Universidad de Buenos Aires, Pabell\'on 1, Ciudad Universitaria, Ciudad Aut\'onoma de Buenos Aires (1428), Argentina and IFIBA - CONICET}
\begin{abstract}
The individual-based models constitute a set of widely implemented tools to analyze the incidence of individuals heterogeneities in the spread of an infectious disease. 
In this work we focus our attention on human contacts heterogeneities through two of the main individual-based models: mobile agents and complex networks models. 
We introduce a novel mobile agents model in which individuals make displacements with sizes according to a truncated power-law distribution based on empirical evidence about human mobility. 
Besides, we present a procedure to obtain an equivalent weighted contact network from the previous mobile agents model, where the weights of the links are interpreted as contact probabilities.
From the topological analysis of the equivalent contact networks we show that small world characteristics are related with truncated power-law distribution for agent displacements. 
Finally, we show the equivalence between both approaches through some numerical experiments for the spread of an infectious disease.
  
\end{abstract}
\maketitle

\begin{section}{Introduction}\label{sec1}

The spread of a disease in human populations had initially been studied by means of deterministic and stochastic compartmental models in which individuals
 are classified according to their status as susceptible (S), infective (I) or refractory (R) \cite{McKendrick}. The original approach implicitly introduces the assumption of totally mixed
 population, where all the agents can interact each other, that allows a simple mathematical formulation \cite{Anderson,AnderssonBritton}.  
Subsequently, the need to incorporate spatial and population heterogeneities has led to some reformulations \cite{LevinDurrett,Hethcote}. 

The path recently followed to describe the social interactions that lead to infectious contacts is the adoption of complex networks as a substrate on which an infectious disease spreads \cite{NewmanSIAM,Latora2,Albert}.
The introduction of complex networks as elements of description has mainly provided the incorporation and study of contacts heterogeneities. 
Thus, in this context, each agent interacts with a finite number of neighbors, which in turn varies from agent to agent.
Many of the efforts have been concentrated on the study of the impact of specific network topological features in the spread of an infectious disease \cite{Keeling,Boguna1,Barthelemy,Moore,Newman1}. 
In particular, the incidence of degree distribution has been extensively analyzed with respect to occurrence probability and total size of an outbreak, especially in some networks with power law degree distribution
for which has been shown the absence of epidemic threshold for susceptible-infective-susceptible dynamic (SIS) \cite{PastorSatorras,PastorSatorras2,May,Boguna}. 
As another example, in \cite{KupermanAbramson} the authors have found a transition from endemic to epidemic oscillatory behavior for SIRS model in networks with Small World property.

Recent years have yielded notable advances in the study of human displacements driven by new possibilities offered by mobile technology systems.  
In a recent paper \cite{Gonzalez} the authors analyze the pattern of movement of a sample of cell phone users finding a kind of truncated L\'evy flight trajectories with a high degree of regularity, contradicting earlier models that described the human displacements from an uniform random walk.
Although L\'evy flight trajectories have been previously suggested to describe the human mobility \cite{Brockmann},
 this does not necessarily imply that each individual must follow a pure Levy flight displacement distribution, as pointed in \cite{Gonzalez}. 
It should be noted that L\'evy flight displacements have been extensively observed in nature and, for example, are believed to constitute the optimal food search strategy for some animals \cite{Viswana1,Viswana2}.

These findings concerning human mobility allow radical improvements on another widely extended modeling tool like mobile agents (MA) models. Its aim is, essentially, seeking to replicate human movement
 and, for example, its impact on the spread of an infectious disease transmitted by closeness or direct contact. In \cite{Eubank} the authors perform realistic simulations
of infectious disease transmission in the city of Portland, Oregon, USA. Another MA model has been proposed to reproduce the topological features of some networks of acquaintances \cite{Gonzalez1}.

On the other hand, while the scientific community has begun to shed light on the impact of different topological features of networks in the spread of human infectious diseases, 
little is known about the origins of such features in social contact networks.

In this paper we consider the spread of a close-contact infectious disease with SIR dynamic through a population of mobile agents.
With this aim we propose a set of MA models considering different movement patterns for the intervening agents.
In particular we consider two alternatives: one where each agent performs an uniform random walk, and the other one that includes recent findings
about human mobility, with displacements sizes that follow a truncated power law distribution as discussed in \cite{Gonzalez}.
On the other hand, we propose an alternative and, as shown later, equivalent complex network-based approach.
For this, we show a method to build an equivalent weighted contact network where the weight of each link indicates its existence probability.

After introducing both approaches we focus on drawing a correspondence between the displacement pattern of agents and topological characteristics of the associated contacts networks.
From this analysis we find that the Small World characteristic of the equivalent contact networks is related to the agents displacements whose sizes follow a truncated power law distribution. 
Finally, we analyze the impact of periodicity in agent displacements in order to reproduce recurrent behavior \cite{Barabasi1}. 

The rest of this paper is organized as follows. In Section \ref{sec2} we present the details of our MA model as well as the criterions of contact and infection transmission between agents.
In Section \ref{sec3} we introduce the procedure to build an equivalent weighted contact network and propose alternative quantities to describe its topological characteristics.
Section \ref{sec4} presents some numerical examples that show the equivalence between MA models and its associated contact networks. Also we analyze the topological characteristics of the equivalent 
contact networks and discuss the link between usual topological features of some social networks and agents mobility pattern.  
Finally, Sec. \ref{sec5} summarizes our conclusions and suggestions for future works.
  
\end{section}

\begin{section}{Mobile agents model}\label{sec2}

Here we propose a stochastic MA model for the transmission of an infectious disease applied on a population of $N$ mobile agents, disposed on a square shaped cell of linear size $L$.
Initially, agents are ordered on a regular square lattice with distance $a=L/(\sqrt{N}+1)$ between nearest neighbors. The position for each agent in this initial configuration will be named \textit{\textquotedblleft home\textquotedblright} from now.
 
In particular, we consider two alternative patterns of motion for individual agents: 1) displacements follow truncated power law distribution or 2) uniform random displacements. 
Regarding scenario 1), recent studies from the trajectories of a significant number of cell phone users have shown that human mobility has a high degree of regularity \cite{Barabasi1}. It was noted that the distribution of displacements of the analyzed sample can be well approximated
by a truncated power law \cite{Gonzalez}:
\begin{equation}\label{ec1}
 P(\Delta r)\propto(\Delta r+\Delta r_0)^{-\beta}\exp(-\Delta r/\kappa)
\end{equation}
where $\beta=1.75 \pm0.15$, $\Delta r_0$ is a characteristic displacement length and $\kappa$ a cutoff value. Moreover, in \cite{Gonzalez} the authors show that the characteristic size of the displacements of each individual to a time $t$, namely the radius of gyration $r_g$, also follows a truncated power law:
\begin{equation}\label{ec2}
 P(r_g)\propto(r_g+r_g^0)^{-\beta_r}\exp(-r_g/\kappa_r)
\end{equation}
with $\beta_r=1.65\pm0.15$ \cite{Gonzalez}. Under these assumptions, they conclude that all agents are characterized by a single distribution of step size independent of their $r_g$. Then, the distribution of displacements for an agent $P_{r_g}(\Delta r)$ can be represented as $P_{r_g}(\Delta r)=r_g^{-\alpha}F(\Delta r/r_g)$, where $F$ is a scaling function satisfying 
$F(x)\sim x^{-\alpha}$ for $x<1$ and a rapid decreasing for $x\gg 1$.

In this scenario, we assign to each agent $i=1,...,N$ a radius of gyration $r_g^i(t)$ from Eq. (\ref{ec2}) with $\kappa=L=1$ and $r_g^0=a$. Then, the agents perform
displacements with randomly chosen direction whose sizes are distributed according to a power law so that its radius of gyration $r_g^i(t)$ corresponds to the values previously assigned.\\

Our second alternative for agent displacements is based on the usual simplified representation of individuals as \textit{random walkers}. 
As in the previous case each agent chooses a random direction, but now the displacement sizes are uniformly distributed within a given interval.
We show examples from both scenarios in Section \ref{sec4}.\\ 

\subsection{Displacement probability}
As is well known, some people make more movements per day than other: it is highly probable that 
a postman makes more displacement a day than an office worker. In order to incorporate these heterogeneities in our model we have assigned each agent a displacement probability
per step through a predetermined probability distribution. We focused on two usual distributions: i) Poissonian and ii) Power-Law type distribution $P_i(x)\sim x^{-\gamma}$, where $x$ represents the number of movements per day made by agent $i$. For example, if $P_i = 1$, the agent $i$ will make a displacement for every step of the temporal evolution.
As regards the time evolution, we consider a discrete evolution taking $1$ hour as our time unit, in other words, every step of our algorithm represents an evolution over time of $1$ hour. Then, the maximum number of movements per day for a given agent will be $x=24$. Finally, at the end of each day the agents return home with constant return probability $P_{ret} = 0.9$, in order to consider some trips more prolonged in time too.\\

In order to keep constant the number of agents in the cell we establish periodic boundary conditions for the movements but not for interactions, i.e,
 we do not consider periodic boundary conditions to calculate distance between agents. This means that when an agent leaves the cell through one of its sides, 
he will be re-entering by the opposite one, however two agents located in opposite sides do not interact.\\

\subsection{Periodic movement pattern}
As human beings we have acquired habits which lead us to reiterate our movements with some frequency, for example: some of us take the kids to school in the morning, then go to the office, less frequently go to gym and finally return home. 
It is possible that sometimes we perform specific non-routine activities but essentially our trajectories are predictable \cite{Barabasi1}. Therefore, in order to reproduce this behavior, our model incorporates an independent periodic motion pattern for each agent. This means that each agent $i=1,...,N$ periodically repeats a given sequence of $q_i$ positions $\mathbf{r^i_1},\mathbf{r^i_2},
\mathbf{r^i_3},...,\mathbf{r^i_{q_i}}$ consistent with previously established scenarios. This is achieved by assigning to each agent $i=1,...,N$ a period $t_i$. In our particular case, the set $\{t_i\}_{i=1,...,N}$ is uniformly distributed in $[1,T_p]$, where $T_p$ is the maximum allowed period. 

\subsection{Infectious disease transmission}\label{sec2b}

Now that we have detailed the features of our mobile agents model, we will focus on the transmission process of an infectious disease.
We study the spread of a general and schematic disease corresponding to Susceptible-Infectious-Recovered (SIR) dynamic.
The transmission occurs by close proximity through direct or indirect contact from infected to susceptible agents, as for influenza, measles or chickenpox,
although we will not focus on the specifics features of any given disease. Here we want to present a general framework to eventually incorporate the specificities of a given infectious disease transmitted by close contact.\\

At this stage it is necessary to establish a proximity criterion between agents that leads to the transmission of an infectious disease. In this model we consider that physical contact is not the only way to produce contagion, but just enough that
 a susceptible agent is near an infected one, sharing a given physical space.
This assumption is based on the transmission processes of some infectious diseases, like influenza, which can be transmitted by means of Fl\"ugge's droplets emitted by a sneeze of an infected individual.

We start the process with a single infected individual located at the center of the square cell. After updating the position of all agents to perform one evolution step, a contact radius $r_c$ is established around its positions.
Thus we say that every pair of agents at a distance less than $r_c$ are \textit{in contact} or \textit{linked}.
Then, the probability that a susceptible agent $i$ becomes infected during an evolution step is given by $P_i (z_i)=1-(1-p_c)^{z_i}$, where $z_i$ is the number of infected individuals found at distance less than $r_c$ from agent $i$ and $p_c$ is the
infection probability per contact. Each infected agent remains in that state for $ \tau_i $ steps of time evolution, after which it becomes recovered (R) and then immune to disease. It should be noted that the updating process is performed synchronously.

\end{section}

\begin{section}{Equivalent complex contact networks}\label{sec3}

So far we have described a MA model where the disease transmission is associated with the movement pattern followed by agents. 
This may help us to understand how spatial displacements of people affects the spreading of an infectious disease.

Another alternative path followed in the literature to incorporate spatial heterogeneities in epidemiological modeling is to introduce a complex network as a substrate on which 
a disease spreads. Thus, the network nodes represent individuals, while the edges their contacts or relationships. 
There is abundant literature about infectious disease spreading on contact networks with diverse topological features \cite{Keeling,Boguna1,Barthelemy,Moore,Newman1,PastorSatorras,PastorSatorras2,May,Boguna}. 
However, there are few cases in which it has been shown the correspondence between the studied system and the proposed contacts network.

In this section we will show how an equivalent weighted and undirected contact network can be drawn from the MA model described in previous section.

\subsection{Network construction. Time window}
As stated in our proximity criterion in Section \ref{sec2b}, we say that two agents are in contact when they are at distance less than $r_c$.
In this case, those nodes, say $i$ and $j$, will be joined by a link $l_{ij}$ whose weight $w_{ij}\neq 0$.
The set of nodes representing agents, together with their links are what we call \textit{equivalent contact network}. 
For the purpose of its construction we need to register contacts between agents during a time interval or \textit{time window} $T_W$. 
If all agents have a recurrent behavior or, still better, a periodic motion pattern, a given agent $i$ repeats its movements with period $t_i$, visiting the same sites over and over again.
 
In this case $T_W$ must be long enough to allow all agents to complete their movement patterns, i.e.,
\begin{equation}\label{ec3}
T_W\geq \max_{i=1,...,N}(t_i) 
\end{equation}
However, in Section \ref{sec4} we will show through some examples that the presumption of recurrent behavior in agents movements 
is not a necessary condition for the construction of an equivalent contact network.
It should be noted that for non-recurrent displacement patterns it becomes necessary to consider the infectious period $\tau_i$ as an additional constraint on $T_W$ value, resulting $T_W\ge \tau_i$.

\subsection{Probabilistic interpretation of weights}
Following our consideration about network construction, it is clear that some contacts will be more frequent than others, so that the resultant contact network will be weighted. We know the identities of all contacts for every agent $i$ over $T_W$ and also the number of times they meet each other.
Then, link weight $w_{ij}$ between nodes $i$ and $j$ can be given by the number of contacts per unit time between these nodes during $T_W$, i.e., $w_{ij}=(\text{number of contacts between $i$ and $j$ during $T_W$})/T_W$. (There exist other factors that contribute 
to the heterogeneity of contacts among agents, like the distinction by age group, race or previous history of disease, neglected in this study with respect to contacts frequency.)
However, here we give a probabilistic interpretation of weights.
If we define $n_{ij}$ as the total number of contacts between nodes $i$ and $j$ during $T_W$, then, the elements $w_{ij}$ of weight matrix $\mathbf{W}$
are given by: 

\begin{equation}\label{ec4}
w_{ij}=w_{ji}=\frac{n_{ij}}{24\times T_W}
\end{equation}

During each time evolution step nodes establish contact only with a subset of their neighbors, chosen with probabilities given in Eq. \ref{ec4}.
Thus the actual pattern of contacts makes stochastic changes over time in order to reproduce the  behavior of MA model.
As a result of the described approach we obtain a matrix $\mathbf{W}$ that contains the fundamental features of the equivalent contact network, such as its undirected character evidenced in Eq. \ref{ec4}.   

\subsection{Infection process in the equivalent contact network}
The infection spreading process in terms of the equivalent contact network is as follows:
\renewcommand{\labelitemiii}{}
\begin{itemize}
 \item[1.] We start the spreading process with one infected agent placed at the center of the cell, as we do for MA model. Each agent $i=1,...,N$ has an assigned counter $t_c(i)$ that defines its state as usual for SIR infectious diseases:
           \begin{equation}
            \begin{array}{cc}
             t_c(i)=0 & \text{agent $i$ is S}\\
             1\leq t_c(i)\leq \tau_i &\text{agent $i$ is I}\\
             t_c(i)\geq \tau_i &\text{agent $i$ is R}
            \end{array}
           \end{equation}
  
 \item[2.] Each susceptible node makes contact with a given set from its neighborhood according to contact probabilities established by matrix $\mathbf{W}$.
 \item[3.] Now we apply the same infection criteria as for MA model: Suppose that a given node $i$ makes contact with $n_i^*$ nodes from its neighborhood, $z_i$ of which are infected. Then, $i$ will become infected with probability $P_i (z_i)=1-(1-p_c)^{z_i}$.
 \item[4.] We perform the contagion criteria synchronously over all S agents.
 \item[5.] The counters of all infected agents are actualized: $t_c(i)=t_c(i)+1$ if agent $i$ is in infected state (I).
 \item[6.] The states of the infected agents are actualized so that each infected agent $i$ remains in that state if $t_c(i)\leq \tau_i$ or it makes a transition to recovered state (R) if $t_c(i)>\tau_i$. This completes one evolution step.
\end{itemize}

Figure \ref{fig1} shows an example of an equivalent contact network obtained for $N=10̣̣̣^4$ mobile agents following a truncated power-law displacements distribution over a time window of $T_W=7$ days.

\begin{figure}
 \centering
 \includegraphics[width=90mm,height=100mm,keepaspectratio=true]{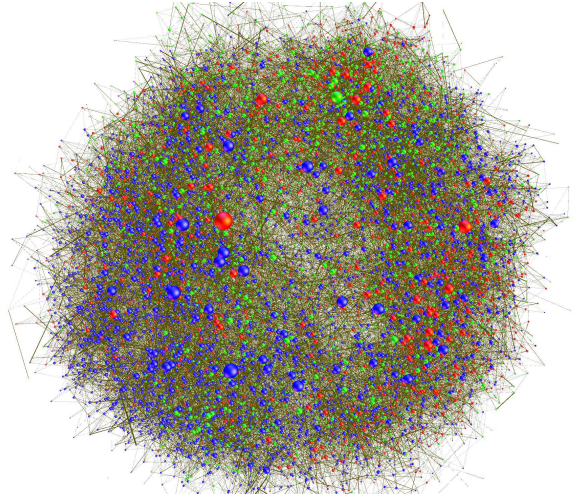}
 \caption{(Color online) Equivalent contact network obtained from a MA model with $N=10^4$ agents following a truncated power-law displacements distribution. Each node has been drawn with a size proportional to its degree
and with a color representing its state: susceptible (green), infected (red) or refractory (blue).}
 \label{fig1}
\end{figure}

Now we present a reformulation of two classical topological quantities, i.e., 
average path length ($\ell$) and clustering coefficient  ($C$), in order to characterize this weighted contact networks.

\subsection{Average path length and clustering coefficient in weighted contact network}\label{sec3b}

One of the quantities that characterize the topology of complex networks is the average path length. It can be defined as the average minimum distance between any pair of nodes \cite{NewmanSIAM,Latora2} and, from now, we call it the \textit{topological} average path length ($\ell$). 
When we refer to distance in this case we mean the minimum number of links that must be traversed to communicate a given pair of nodes.
There exist several algorithms in the literature to find the shortest-path in non-weighted networks, some examples are Dijkstra's algorithm, Breadth-first search and 
Floyd-Warshall algorithm. In addition, some of them can be generalized for weighted networks.

Generally, for ordinary weighted networks, each link has an associated weight corresponding, for example, to its ability to carry information or, as in the problem studied here, spreading a disease. 
According to our probabilistic interpretation, each weight indicates the probability of link existence. Then, a given path leading from node $i$ to $j$ through the links
 $l_{i,n_1}$, $l_{n_1,n_2}$, $l_{n_2,n_3}$, ... , $l_{n_m,j}$ will have existence probability $P_{i\rightarrow j}=w_{in_1}w_{n_1n_2}w_{n_2n_3}...w_{n_mj}$. 
This is why it is necessary an additional definition of average path length for this probabilistic interpretation of weights. The alternative definition should include both the distance $d^k_{ij}$ between nodes $i$ and $j$ through path $k$ as the existence probability of that path. Then, we call it the \textit{probabilistic} average path length ($\ell_p$).

The probability $P^k_{i\rightarrow j}$ can be interpreted as a measure of the persistence of the path $k$ that connects node $i$ to $j$, so that the probabilistic shortest path is the one that maximizes the probability $P^k_{i\rightarrow j}$.
In this sense, we say that $d_ {ij}=d^L_ {ij}$ is the minimum distance between $i$ and $j$ if:  
\begin{equation}\label{ec5}
 P^L_{i\rightarrow j}=\max_{k\in \{all\:paths\:i\rightarrow j\}}\left(P^k_{i\rightarrow j}\right)
\end{equation}
where $k$ runs through all paths connecting nodes $i$ and $j$ and we call $L$ to the probabilistic shortest path that connect them. Finally, the probabilistic average path length ($\ell_p$) will be the average of $d_{ij}^L$ between all possible pair of nodes in the network:
\begin{equation}\label{ec8}
 \ell_p=\frac{1}{N(N-1)}\sum_{i,j=1}^N d_{ij}^L
\end{equation}

Another quantity used to characterize network topology is the average clustering coefficient $C$. It measures the likelihood that two neighbors of a node $i$ are associated among themselves \cite{NewmanSIAM,Latora2}. The local version of clustering coefficient $c_i$
can be defined for every node $i$ as the fraction of connected neighbors of $i$, so that the average clustering coefficient is 
\begin{equation}\label{ec6}
C=\frac{1}{N}\sum_i c_i \text{  .}  
\end{equation} 

If we call $n(i)$ the set of $k_i$ neighbors of $ i $ and we define $w_i=\sum_{j\in n(i)}w_{ij}$, then the local clustering coefficient can be written as:
\begin{equation}\label{ec7}
 c_i=\frac{1}{w_i(k_i-1)}\sum_{j,k\in n(i)} w_{jk} \text{  .}
\end{equation}

The definition of Eq. \ref{ec7} applies to both weighted and non-weighted networks since when $w_{ij}=1$ $\forall i,j=1,...,N$ usual definition for non-weighted networks is recovered.

\end{section}

\begin{section}{Numerical simulations}\label{sec4}
 We consider a mobile agent model with
 $N=10^4$ agents within a square cell of side $L=1$ (in adimensional units). 
We fix the contact radius $r_c$ equal to half the distance between nearest neighbors in the initial square lattice configuration, i.e. $r_c=a/2=L/(2(\sqrt{N}+1))$.
Actually, the value of $r_c$ will depend on the disease characteristics but, as already mentioned, we are not concerned here with any particular disease.

The displacement probability for each agent was assigned from a Poisson distribution $P(x)=\frac{\lambda^x}{x!}e^{-\lambda}$ with $\lambda=2.1$.

Each agent performs a periodic motion with step size according to the truncated power law distribution described in Sec. \ref{sec2}. We will refer to this model as \textit{periodic truncated power law displacements} (PTPL). The periods of movement $\{t_i\}_{i=1,...,N}$ are chosen with uniform probability from $[1,7]$, i.e. from one day to a week.

\subsection{Time window election and topological properties of contact networks}\label{sec4b}
Now we will construct an equivalent contact network from PTPL model.
For this we need to determine an optimal time window size $T_W$.  
Common sense dictates that the larger $T_W$ the better will be the description given by the equivalent contact network.
Clearly there is a practical compromise, so it is necessary to establish a lower bound for $T_W$ so that the topological characteristics 
of the equivalent contacts network are not substantially modified.

\begin{figure}
 \centering
 \includegraphics[width=80mm,height=100mm,keepaspectratio=true]{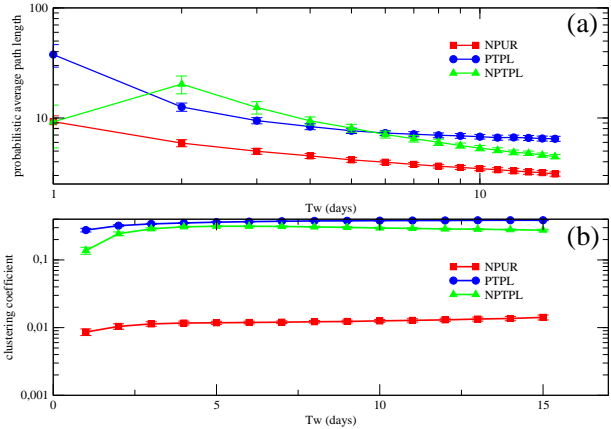}
 \caption{(Color online) Contact network topology characterization for NPUR (red squares), PTPL (blue circles) and NPTPL (green triangles) variants of our MA model. a) Average probabilistic path length versus time window size $T_W$. b) Clustering coefficient for different $T_W$. 
Each point correspond to average over $1000$ contact networks derived from independent realizations of each variant of MA model.}
 \label{fig2}
\end{figure}

As can be seen in Fig. \ref{fig2} for PTPL, both $\ell_p$ and $C$ approach their asymptotic values for $T_W\geq7$, consistent with the maximum period of our mobile agents.
It should be noted that $\ell$ and $\ell_p$ are approximately equal for the studied cases and therefore we do not show both results.
The equivalent contact network obtained for $T_W=7$ has average degree $\langle k\rangle=8.3\pm 0.6$ and Small World topological characteristics with $\ell=\ell_p=7.12\pm 0.38$ and $C=0.374\pm 0.004$.

It is important to note here that the Small-World features arise naturally from the imposition of a movement pattern for agents, 
with displacements sizes according to a truncated power law distribution. 

On the other side, the incorporation of periodicity in 
agent movements limits the spatial locations available to each agent, reducing the time window necessary to obtain a correct sampling of possible contacts between them.
To verify this last assertion we repeat the analysis above but now for two alternative scenarios: i) agents following  \textit{non-periodic truncated power law} (NPTPL) displacements, and ii) agents carrying out \textit{non-periodic uniformly distributed random} (NPUR) displacements. 

\begin{figure}
 \centering
 \includegraphics[width=80 mm,height=100 mm,keepaspectratio=true]{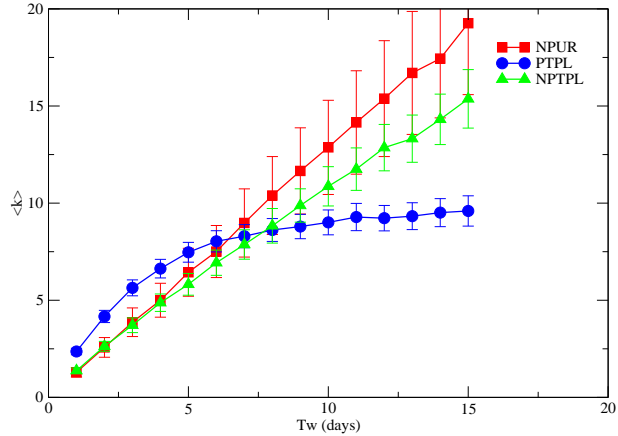}
 \caption{(Color online) Average degree $\langle k\rangle$ in terms of time window size $T_W$ for equivalent contact networks obtained from NPUR (red squares), PTPL (blue circles) and NPTPL (green triangles) mobile agents displacement models. Again, each point in the graphic corresponds to average over $1000$ independent realizations of each variant of MA model.}
 \label{fig3}
\end{figure}

As shown in Fig. \ref{fig2}, NPUR model has slower convergence to the asymptotic regime compared with PTPL model, in particular decaying as a power law. 
This behavior is consistent with the linear dependence between average degree and $T_W$ observed for NPUR model in Fig. \ref{fig3}. 
In contrast, when agents follow a periodic movement pattern as in PTPL, the curve representing the evolution of $\langle k\rangle(T_W)$ also tends to an asymptotic regime.
The NPTPL model shows an intermediate behavior between the previous ones. Clearly, when $T_W = 1$ is difficult to distinguish between a model with agents making uniform 
random displacements (NPUR) and another whose agents follow a truncated power-law step size distribution without any periodicity (NPTPL).
By increasing $T_W$ the contact network associated to NPTPL model begins to show the Small World characteristics also observed for PTPL displacements model (see Fig. \ref{fig2}).

On the other hand, setting $T_W = 15$ for NPUR model we obtain an equivalent contact network with $\ell=\ell_p=3.1\pm 0.1$ and $C=0.014\pm0.001$, characteristics consistent with Random Networks.

Despite the foregoing, in Section \ref{sec5} we show that the disease spreading dynamic for the NPUR model can also be reproduced with very good matching through its equivalent contact network.

\begin{figure}
 \centering
 \includegraphics[width=80 mm,height=100 mm,bb=0 0 403 279,keepaspectratio=true]{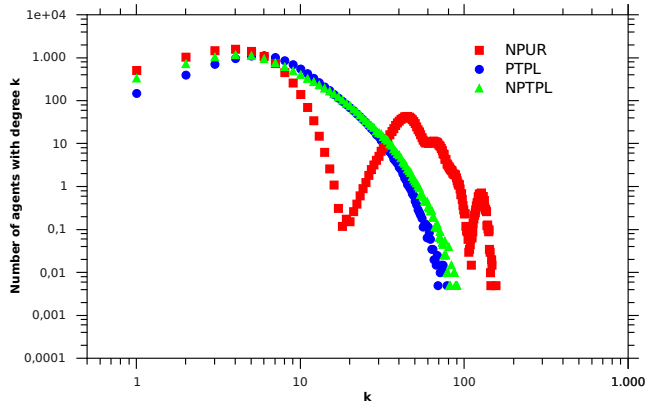}
 \caption{(Color online) Degree distribution for contact networks derived from NPUR (red squares), PTPL (blue circles) and NPTPL (green triangles) mobile agents models. In all cases we consider $T_W=7$ and we average over $1000$ independent realizations of each variant of our MA model.}
 \label{fig4}
\end{figure}

Figure \ref{fig4} shows the degree distributions for PTPL, NPTPL and NPUR models, all for $T_W=7$ days.
For PTPL the degree distribution initially behaves like a Poisson distribution but then decays more slowly, following a power-law with exponent $\tau=-3.2$ for $10 <k <35$.
This heavy-tailed degree distribution has been previously reported for urban social contact networks \cite{Eubank,Onnela}. 

In contrast, degree distribution for NPUR shows an irregular behavior resulting in an increased frequency of high-degree agents.
NPTPL resembles the behavior of NPUR for $1<k<5$ but for $k\geq5$ it behaves like PTPL.
     
\subsection{Infectious disease spreading. Comparison between Mobile-agents model and its equivalent contact- network.}
We will now analyze the spread of an infectious disease with SIR dynamic in a population of PTPL mobile agents, in order to compare its results with those obtained from the 
equivalent contact network approach. We select a fixed infectious period $\tau_i=4$ for all agents and a time window $T_W=7$, that satisfies $T_W\geq \max_{i=1,...,N}(t_i)$ and $T_W\geq \tau_i$.
Figure \ref{fig5} shows the time evolution of total infected agents for PTPL mobile agents model and its equivalent contact network.
It can be seen that both approaches show a very good agreement. 
\begin{figure}
 \centering
 \includegraphics[width=85mm,height=95mm,keepaspectratio=true]{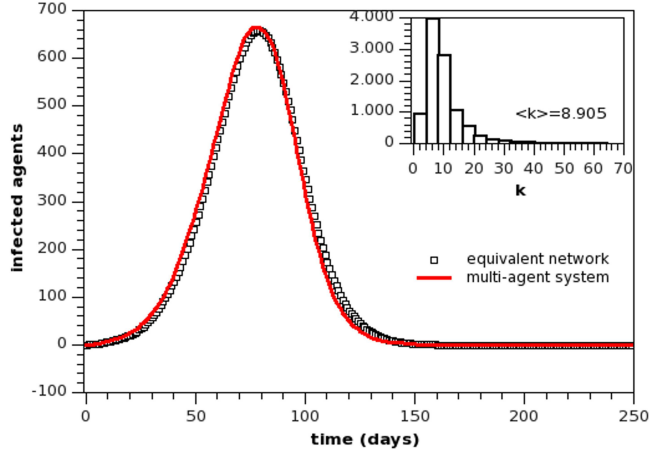}
 \caption{Dynamical comparison between the MA model and its equivalent contact network approach for the spread of an infectious disease with SIR dynamic. The time evolution of total infected individuals for PTPL mobile agents model is represented by the solid line in the figure, 
while the squares correspond to the results for its equivalent contact network. All the results were obtained from the average over $10000$ independent realizations of the disease spreading process on a single spatial realization of the PTPL mobile agents model and its equivalent contact network.
The inset shows the degree distribution for the contact network with average degree $\langle k\rangle=8.905$, obtained from the imposition of a poissonian displacement probability for the agents in MA model.}
 \label{fig5}
\end{figure}

Equivalently, in Figure \ref{fig6} we compare the total number of infected agents for NPUR mobile agents model
with results for equivalent contact networks corresponding to different time window sizes.
The good agreement observed for $T_W\geq 6$ in Fig. \ref{fig6} can be understood remembering the constraint $T_W\geq \tau_i$ and that, in the context of NPUR model, each mobile agent makes an uniform random walk, thereby increasing its number of contacts with time, 
thus improving the sampling performance. 

\begin{figure}
 \centering
 \includegraphics[width=82mm,height=100mm,keepaspectratio=true]{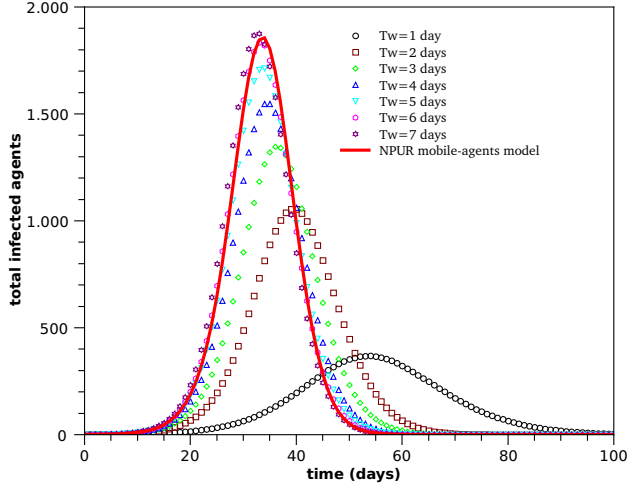}
 \caption{Comparison of the total infected agents evolution for a set of equivalent contacts networks. The sucession of equivalent contact networks has been obtained by increasing the sampling time $T_W$ from $1$ to $7$ days over a single spatial realization of the NPUR mobile agents model.
The solid line shows the time evolution of infected individuals for NPUR mobile agents model. Again, all the results correspond to averages over $10000$ independent realizations of the disease spreading process.}
 \label{fig6}
\end{figure}

So far we have arbitrarily assumed that the displacement probability, i.e. the probability that an agent makes a 
displacement, follows a Poisson distribution. Now we show that the observed agreement 
between both approaches is not constrained to this condition by imposing a power-law displacement distribution $P(x)\sim (x-x_0)^{-\gamma}\exp(-x/k_0)$ with exponential cutoff.
All parameters were selected so that the average degree is close to that obtained for the example shown in Fig. \ref{fig5}.
The equivalent contact network also shows Small-World topological characteristics with $\ell\approxeq\ell_p=6.977$, $C=0.383$ and average degree $\langle k\rangle=8.324$.
Again, we show in Fig. \ref{fig7} the good agreement between PTPL mobile agents model, in this case with power-law displacement probability, and its equivalent
contact network, for temporal evolution of infected agents. 
 
\begin{figure}
 \centering
 \includegraphics[width=80mm,height=100mm,keepaspectratio=true]{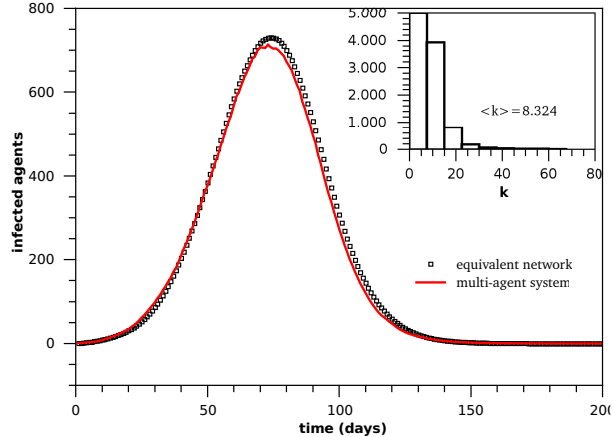}
 \caption{Time evolution of the infected individuals for PTPL mobile agents model with power law displacement probability (solid line) and its equivalent contact network (squares). The inset in the figure shows the degree distribution of the equivalent contact network, which now resembles a power-law behavior.}
 \label{fig7}
\end{figure}

\end{section}

\begin{section}{Summary and conclusions}\label{sec5}
In this paper we introduce a novel MA model in which agents make displacements according to a particular truncated power-law distribution consistent with recent findings related to human mobility \cite{Gonzalez,Brockmann}. 
We have developed a procedure to build an equivalent weighted contact network by recording contacts through a proximity criteria between mobile agents during a given time window. The weights are interpreted here as contact probabilities and this fact provides stochasticity to the contact network. 
Given the probabilistic nature of the weights we have introduced alternative definitions for the topological network descriptors. In particular we have defined a probabilistic average path length which results equivalent to the usual topological average path length in the studied cases.
We want to note that the analyzed contact networks are not necessarily friendship networks or coworkers networks but just are networks of people who have shared the same spatial region for a period of time. However, contact networks can also provide information about friend or work relationships among 
individuals through the weights of the links that bind them.   

We have analyzed three different agents movement patterns for which we have studied the topological features of its equivalent contact networks. Our results provide evidence that relates the truncated power-law displacement pattern of agents with Small World characteristic of its equivalent social contact networks.
The former correspondence can be explained by considering that this kind of movement pattern increases the encounters probability among agents whose homes are close with respect to those who are far away. In this way, there will be tightly interconnected sets of agents, contributing to increase the average clustering coefficient, 
which are linked to others by sporadical long displacements that act as shortcuts. We are currently addressing the study of community structure formation from different displacement patterns and its impact on spreading dynamic in complex networks.
Another important factor to consider, related to the previous one, is the distribution of homes in the cell. In this paper we have considered a regular and uniform distribution of homes in order to reduce the number of parameters involved in the problem. 
However, this is an appropriate assumption only for some urban areas, so it is necessary to extend our MA models to study the impact of demographic heterogeneities.  

The contacts networks obtained from PTPL model with regular reincidence in agent positions reached its asymptotic regime, with respect to topological variables, faster than the other two displacement models without reincidence. This assumption of regularity in human mobility has recently received empirical support \cite{Barabasi1},  
allowing us to optimize the time of sampling that leads to equivalent contact networks. However, we have shown through an example for the spread of a SIR infectious disease with finite infectious period, that a MA model with non-recurrent movement pattern as NPUR model shows good agreement 
with the results of its equivalent contact network.  

Finally, we have shown the equivalence between mobile agents and contact networks models for the spreading of a particular kind of infectious diseases with SIR dynamic and transmitted by proximity.

\end{section}

\acknowledgements{A.D.M. is grateful to Universidad de Buenos Aires for financial support through its postgraduate fellowship program. A.D.M and C.O.D acknowledge financial support from Universidad de Buenos Aires under project No. 20020090200476.}

\end{document}